# Parametrized KAM Theorem for Differentiable Hamiltonian Vector Fields without Action-Angle Variables

Wu-hwan Jong[1] & Jin-chol Paek[2]

1) *Information Technology Institute, Kim Il Sung University, D.P.R.K*
2) *Department of Mathematics, Kim Il Sung University, D.P.R.K*
Email: wuhuanjong@yahoo.com

## Abstract

We proved a parametrized KAM theorem in Hamiltonian system which has differentiable Hamiltonian without action-angle coordinates. It is a generalization of the result of [Llave et al. 2005] that deals with real analytic Hamiltonians.

## 1. Introduction

In this paper, we have presented a KAM theorem on existence of invariant tori with a Diophantine vector for differentiable Hamiltonian vector fields with parameters. We deal with differentiable Hamiltonian vector field which does not require to be a perturbation of an integrable one or to be written in action-angle variables.

The existence problem of invariant tori for Hamiltonian system is often appeared in various scientific fields ([Celletti-Chierchia 2006], [Fèjoz 2004], [Locatelli-Giorgilli 2007], [Helleman-Kheifets 1985], [Casartelli 1983], [Gidea, 2009] etc).

[Kolmogorov 1954] has proposed the procedure which clarifies the existence of invariant tori for perturbed real analytic Hamiltonian vector field with action-angle variables at first and [Arnold 1963] has given the rigorous proof based on Kolmogorov's procedure. [Moser 1962] has proved the existence of invariant tori for real analytic area-preserving twist mappings on 2-dimensional annulus with action-angle variables and moreover he has relaxed the assumption of alalyticity of the map to $C^{333}$ differentiability. After that [Rüssmann 1970] has relaxed the differentiability condition with the existence of invariant tori in Hamiltonian system to $C^5$ class and [Takens 1971] has clarified that it is not enough for $C^1$ class. Finally [Herman 1986] has clarified that it is enough for $C^3$ class but not to $C^2$-mappings whose second derivatives belong to the Hölder class $C^{1-\delta}$ where $\delta > 0$ is small(see [Broer- Sevryuk 2008], pp.21).

However these results are for the Hamiltonian systems to be a perturbation of an integrable one and to be written in action-angle variables. On the other hands, action-angle variables have singularity at elliptic fixed points or in neighborhood of separatrix and the use of action-angle variables are too restrictive in the case of numerical analysis. Moreover, in many practical applications, we have to consider the system which is not near to integrable one but has approximate invariant tori with sufficiently small error([Llave et al. 2005]). [Llave et al. 2005] proved the existence of invariant tori for real analytic Hamiltonian system which is not a perturbed integrable one or is written in action-angle variables and [Haro-Llave 2004a, 2004b] applied this result in the numerical computation of invariant tori.

The parametrized KAM theory gives an approach to KAM theory of lower dimensional invariant tori in Hamiltonian systems([Broer- Sevryuk 2008], pp. 44). [Broer et al. 1990], [Broer et al. 2007], [Hoo 2005], [Huitema 1988] have got parametrized KAM theorem for Hamiltonian system and [Broer et al. 1990], [Broer et al. 2007], [Ciocci et al. 2005], [Hoo 2005], [Huitema 1988] have got parametrized KAM theorem for dissipative system. Meanwhile [Li-Yi 2002] have proved a parametrized KAM theorem for the generalized Hamiltonian system(Poisson system). [Llave et al. 2005] have established a parametrized KAM theorem for Hamiltonian fields which is not a perturbed integrable one or is written in action-angle variables.

In this paper, we proved a parametrized KAM theorem for differentiable Hamiltonian vector fields which is not perturbed integrable one or is not written in action-angle variables.

## 2. parametrized KAM theorem in the case of real analytic Hamiltonian

Let $n$ be a positive integer number. We let $|z|=\max_{1\le j\le n}|z_j|$ for $z\in\mathbf{C}^n$ and $|A|=\max_{\substack{1\le i\le m\\ i\le j\le n}}|a_{ij}|$ for $m\times n$ matrix $A=(a_{ij})$. Let $U_\rho$ denote the complex closed strip of width $\rho>0$: $U_\rho=\{\theta\in\mathbf{C}^n\ ;\ |\mathrm{Im}\theta|\le\rho\}$. Let $B$ be a compact subset of $\mathbf{R}^n$ which is included in closure of its interior i.e. $B\subset\overline{\mathrm{int}B}$. Given a function $g\in C^m(B)$, for $m\in\mathbf{Z}_+=\mathbf{N}\bigcup\{0\}$ we will denote the $C^m$-norm of $g$ on $B$ by $|g|_{C^m,B}$. Given a 1-periodic function $\psi$, continuous on $n$-dimensional tori $\mathbf{T}^n=\mathbf{R}^n/\mathbf{Z}^n$, we denote the average of $\psi$ on $\mathbf{T}^n$ by $<\psi>=\int_{\mathbf{T}^n}\psi(\theta)d\theta$.

**Definition 1**. Given $\gamma>0$ and $\sigma>n-1$, we define $D(\gamma,\ \sigma)$ as the set of frequency vectors $\omega=(\omega_1,\ \cdots,\ \omega_n)\in\mathbf{R}^n$ satisfying the *Diophantine condition*:

$$|k\cdot\omega|\ge\gamma|k|_1^{-\sigma},\ \forall k\in\mathbf{Z}^n-\{0\}$$

where $|k|_1=|k_1|+\cdots+|k_n|$.

**Definition 2.** If $V$ is a open subset of topological space of $X$ then we denote this fact by $V\overset{\circ}{\subset}X$. We will assume that $\mathbf{U}$ is either $\mathbf{T}^n\times U$ with $U\overset{\circ}{\subset}\mathbf{R}^n$ or $\mathbf{U}\overset{\circ}{\subset}\mathbf{R}^{2n}$. Let $Q$ is a subset of $\mathbf{R}^d$ $(d\in\mathbf{N})$ which is included in closure of its interior. We suppose that $H:\mathbf{U}\times Q\to\mathbf{R}$ satisfy following two conditions:

1) for each $x\in\mathbf{U}$ the function $H(x,\ \cdot):Q\to\mathbf{R}$ is $C^2$,

2) for each $\lambda\in Q$ the function $H_\lambda:=H(\cdot,\ \lambda):\mathbf{U}\to\mathbf{R}$ is real analytic Hamiltonian function.

Then we call $H=H(x,\ \lambda)$ as $d$*-parametric family of Hamiltonian functions*.

We denote *n*-dimensional unit matrix as $I$. Let $J=\begin{pmatrix}O & I\\ -I & O\end{pmatrix}$. We will consider the first-order partial differential equation

$$\partial_\omega K(\theta)=J\nabla H_\lambda(K(\theta)) \tag{1}$$

for $2n$-parametric family of Hamiltonian functions $H:\mathbf{U}\times Q\to\mathbf{R}$, where unknown is $K:\mathbf{T}^n\to\mathbf{U}$. We call equation (1) as *Hamiltonian invariant torus equation*. In this place, $\omega\in D(\gamma,\ \sigma)\subset\mathbf{R}^n$ and $\partial_\omega$ is the derivative in direction $\omega$:

$$\partial_\omega K=\sum_{i=1}^{n}\omega_i\frac{\partial}{\partial\theta_i}K\ .$$

**Definition 3.** Suppose that $H:\mathbf{U}\times Q\to\mathbf{R}$ is a $2n$-parametric family of Hamiltonian functions and $\omega\in D(\gamma,\ \sigma)\subset\mathbf{R}^n$. $\Pi_\rho$ will denote the Banach space of maps $K:U_\rho\to\mathbf{U}$ which are 1-periodic in all its variables, real analytic on the interior of $U_\rho$ and continuous on the boundary of $U_\rho$ with norm $\|K\|_\rho=\sup_{\theta\in U_\rho}|K(\theta)|$. Let $\lambda_0\in Q$. We suppose that $K\in\Pi_\rho$ satisfy following two conditions:

1) There exists a $n\times n$ matrix-valued function $N(\theta)$ satisfying

$$N(\theta)(DK(\theta)^T DK(\theta))=I\ ,$$

2) $<\Lambda_0>$ is invertible with

$$\Lambda_0(\theta)=\begin{pmatrix} N_0(\theta)DK_0(\theta)^T J\left(\dfrac{\partial}{\partial\lambda}\nabla H_\lambda(K_0(\theta))\Big|_{\lambda=\lambda_0}\right)\\ DK_0(\theta)^T J\left(\dfrac{\partial}{\partial\lambda}\nabla H_\lambda(K_0(\theta))\Big|_{\lambda=\lambda_0}\right)\end{pmatrix}.$$

Then we call $K$ is *non-degenerate*.

**Theorem.** (Theorem 3 in [Llave et al. 2005]) Let $\omega\in D(\gamma,\ \sigma)$. Assume that $K_0\in\Pi_\rho$ is non-degenerate. Assume that $H:\mathbf{U}\times Q\to\mathbf{R}$ is a $2n$-parametric family of Hamiltonian functions and there exist $r>0$ such that for any $\lambda\in Q$, $H_\lambda$ is can be holomorphically extended to $r$-neighbor-hood of the image of $U_\rho$ under $K_0$:

$$B_r=B_r(K_0)=\{z\in\mathbf{R}^n\ ;\ \inf_{\theta\in U_\rho}|z-K_0(\theta)|<r\}.$$

Let's suppose that $Q$ is a compact subset of $\mathbf{R}^n$ which is included in closure of its interior and involve $r$-neighborhood of $\lambda_0\in\mathbf{R}^{2n}$. Define the error function for $\lambda_0$, $K_0$ by

$$e_0(\theta)=J\nabla H_{\lambda_0}(K_0(\theta))-\partial_\omega K_0(\theta)).$$

Let $\delta_0=\min(1,\ \rho/12)$.

Then there exists a constant $c>0$, depending on $\sigma$, $n$, $r$, $\rho$, $|H|_{C^3,B_r\times Q}$, $\|DK_0\|_\rho$, $\|N_0\|_\rho$, $|<\Lambda_0>^{-1}|$ such that if

$$c\gamma^{-4}\delta_0^{-4\sigma}\|e_0\|_\rho<1 \qquad (2)$$

$$c\gamma^{-2}\delta_0^{-2\sigma}\|e_0\|_\rho<r \qquad (3)$$

then there exist $\lambda_\infty\in Q$ and $K_\infty\in\Pi_{\rho/2}$ which satisfies the non-degenerate conditions such that

$$\partial_\omega K_\infty(\theta)=J\nabla H_{\lambda_\infty}(K_\infty(\theta)).$$

Moreover $\lambda_\infty$ and $K_\infty$ satisfy

$$\|K_\infty-K_0\|_{\rho/2}\le r, \qquad (4)$$

$$|\lambda_\infty-\lambda_0|<r. \qquad (5)$$

and $\|DK_\infty\|_{\rho/2}$, $\|N_\infty\|_{\rho/2}$, $|<\Lambda_\infty>^{-1}|$ satisfy

$$\|DK_\infty\|_{\rho/2}\le\|DK_0\|_{\rho_0}+\beta,$$

$$\|N_\infty\|_{\rho/2}\le\|N_0\|_{\rho_0}+\beta,$$

$$|<\Lambda_\infty>^{-1}|\le|<\Lambda_0>^{-1}|+\beta.$$

In this place, $N_\infty$ and $\Lambda_\infty$ are as in definition 3, replacing $K$ with $K_\infty$ and $\lambda$ with $\lambda_\infty$ and $\beta=\gamma^{-2}\delta_0^{2\sigma-1}2^{-4\sigma}$.

**Remark 1.** The dependence of constant $c$ on $|H|_{C^3,\mathrm{B}_r\times Q}$, $\|DK_0\|_\rho$, $\|N_0\|_\rho$, $|<\Lambda_0>^{-1}|$ is polynomial. That is, there exists a polynomial, $\lambda(y_1,\ y_2,\ y_3,\ y_4)$ with positive coefficients depending on $\sigma$, $n$ and such that

$$c=\lambda(|H|_{C^3,\mathrm{B}_r\times Q},\ \|DK_0\|_\rho,\ \|N_0\|_\rho,\ |<\Lambda_0>^{-1}|)$$

(Remark 15 in [Llave et al. 2005]).

## 3. Parameterized KAM theorem in the case of differentiable Hamiltonian

To prove the existence of invariant tori in the case of differentiable Hamiltonian with parameter, we need some approximation propositions.

**Lemma 1.** Let $E_n=[a_1,\ b_1]\times[a_2,\ b_2]\times\cdots\times[a_n,\ b_n]$ and $f\in C^4(E_n)$. Then there exists a sequence of real analytic functions $\{f_k\}$ on $E_n$ such that $|f_k-f|_{C^3,E_n}\to 0,\ k\to\infty$.

**proof**. See [Jong-Paek 2012]. □

**Lemma 2.** Let $l\in\mathbf{N}$. If a sequence of functions $\{f_k(x)\}$ real analytic on $U_{\rho/4^{k-1}}$ satisfies the following inequality

$$\|f_k(x)-f_{k+1}(x)\|_{\rho/4^k}\le A(4^{-l})^k,$$

where $A\ge 0$ is a suitable constant, then $f_k(x)$ converges to certain function $f\in C^1(\mathbf{T}^n)$.

**proof**. See lemma 1 of Chapter 3, Section 7 in [Moser 1966]. □

Our main result is as follows:

**Theorem 1.** Let $\omega \in D(\gamma,\ \sigma)\,(\sigma > n-1)$. Assume that $K_0 \in \Pi_\rho$ and $K_0$ is non-degenerate. Assume that $H : \mathbf{U} \times Q \to \mathbf{R}$ is $C^l$ ($l \ge 4$) and that is can be extended to $B_{3r}(K_0) \times A(Q)$ as $C^l$ class. In this place, the parameter set $Q$ is a compact subset of $\mathbf{R}^n$ which is included in closure of its interior and include $2r$-neighbor of $\lambda_0$ and $A(Q)$ is a some rectangle region involving $Q$. Define the error function $e_0(\theta) = J\nabla H_{\lambda_0}(K_0(\theta)) - \partial_\omega K_0(\theta)$ ($\theta \in U_\rho$) and $\delta_0 = \min(1,\ \rho/12)$.

For certain constant $c > 0$, depending on $\sigma$, $n$, $|H|_{C^3,\ B_{2r} \times Q}$, $\| DK_0 \|_\rho$, $\| N_0 \|_\rho$, $|<\Lambda_0>^{-1}|$,

if error function $e_0$ satisfies (2),(3) then there exist $\lambda_\infty \in Q$ and $C^1$ map $K_\infty : \mathbf{T}^n \to \mathbf{U}$ satisfying

$$\partial_\omega K_\infty(\theta) = J\nabla H_{\lambda_\infty}(K_\infty(\theta)),\ (\theta \in \mathbf{T}^n).$$

**Proof.** We use the following notations:

$$\mu_0 = |H|_{C^3,\ B_{2r} \times Q},\ d_0 = \| DK_0 \|_\rho,\ \nu_0 = \| N_0 \|_\rho,\ \tau_0 = |<\Lambda_0>^{-1}|,$$

$$\beta = \gamma^{-2} \delta_0^{2\sigma-1} \frac{1}{2^{4\sigma} - 2^{2\sigma+1}},\ \mu = \mu_0 + 1,\ d = d_0 + \beta,\ \nu = \nu_0 + \beta,\ \tau = \tau_0 + \beta + 1.$$

Let $c = \lambda(\mu,\ d,\ \nu,\ \tau)$. Since $B_{3r}(K_0)$ is a bounded subset of $\mathbf{R}^n$, there exists a rectangle region $E_{2n}(K_0) = [a_1,\ b_1] \times \cdots \times [a_{2n},\ b_{2n}]$ satisfying

$$B_{3r}(K_0) \subset E_{2n}(K_0).$$

Since $B_{5r/2}(K_0)$ is a open neighborhood of $K_0(U_\rho)$, there exists a function $\varphi \in C^\infty(E_{2n}(K_0))$ such that

$$\varphi(z) = \begin{cases} 1, & z \in K_0(U_\rho) \\ 0, & z \in E_{2n}(K_0) \setminus B_{5r/2}(K_0) \end{cases} \quad \text{and} \quad 0 \le \varphi(z) \le 1,\ \ (z \in E_{2n}(K_0)).$$

Now define the function $\psi : E_{2n}(K_0) \times A(Q) \to \mathbf{R}$ by $\psi := \varphi \circ p_1$, where $p_1 : E_{2n}(K_0) \times A(Q) \to E_{2n}(K_0)$ is projection map. Since projection map is $C^\infty$, $\psi$ is $C^\infty$. Now we extend the function $H$ onto $E_{2n}(K_0) \times A(Q)$ arbitrarily and consider the function $H \cdot \psi : E_{2n}(K_0) \times A(Q) \to \mathbf{R}$. Because $H$ and $\varphi$ are both $C^l$ and $\varphi(z) = 0,\ (z \in (E_{2n}(K_0) \setminus B_{5r/2}(K_0)) \times A(Q))$, $H \cdot \psi$ is $C^l$ on $(E_{2n}(K_0) \setminus B_{5r/2}(K_0)) \times A(Q)$. Therefore $H \cdot \varphi$ is $C^l$ on $E_{2n}(K_0) \times A(Q)$. And $H \cdot \varphi$ preserves function values of $H$ on $B_{2r}(K_0) \times A(Q)$. So we denote $H \cdot \psi$ as $H$. Then from the lemma 1, there exists a sequence of real analytic functions on $E_{2n}(K_0) \times A(Q)$ which $C^3$-converges to $H$. If we take appropriate subsequence of the sequence then we get a sequence of real analytic functions on $E_{2n}(K_0) \times A(Q)$, $\{H^k\}_{k \ge 1}$ such that

$$|H^k - H^{k+1}|_{C^3, E_{2n}(K_0) \times A(Q)} \le A(4^{-k})^{l+2\sigma}, \tag{6}$$

$$|H^k - H|_{C^3, E_{2n}(K_0) \times A(Q)} \le A(4^{-k})^{l+2\sigma}. \tag{7}$$

where $A \ge 0$ is a constant depending on only $|H|_{C^3, E_{2n}(K_0) \times A(Q)}$.

Because $B_{2r}(K_0) \subset E_{2n}(K_0)$, we obtain that

$$|H^k - H^{k+1}|_{C^3, B_{2r}(K_0)\times A(Q)} \le A(4^{-k})^{l+2\sigma}, \quad |H^k - H|_{C^3, B_{2r}(K_0)\times A(Q)} \le A(4^{-k})^{l+2\sigma}.$$

Let us prove the theorem 1 in two steps.

**First Step.** For the first step, we will prove that for some number $k_0$ and $\lambda_{k_0} \in Q$, there exists a solution $K_{k_0} \in \Pi_{\rho/2}$ of (1) with Hamiltonian $H^{k_0}_{\lambda_{k_0}}$ satisfying

$$\| K_{k_0} - K_0 \|_{\rho/2} \le r,$$

$$|\lambda_{k_0} - \lambda_0| \le r.$$

For this, let us prove that for sufficient large number $k_0$, for some $\lambda_{k_0} \in Q$, the pair $(H^{k_0}_{\lambda_{k_0}}, K_0)$ satisfy the assumptions of the theorem in section 2, i.e. $c_{k_0}$ and $e_{k_0}$ defined by

$$c_{k_0} = \lambda(|H^{k_0}|_{C^3, B_{2r}(K_0)\times Q}, \| DK_0 \|_\rho, \| N_0 \|_\rho, |<\Lambda_0^{k_0}>^{-1}|),$$

$$e_{k_0} = J\nabla H^{k_0}_{\lambda_0}(K_0(\theta)) - \partial_\omega K_0(\theta)$$

satisfy (2) and (3). First we prove that $c_{k_0} < c$. Performing some simple computations, we obtain

$$\Lambda_0^{k_0}(\theta) = \begin{pmatrix} N_0(\theta)DK_0(\theta)^T J\left(\frac{\partial}{\partial\lambda}\nabla H_\lambda^{k_0}(K_0(\theta))\Big|_{\lambda=\lambda_0}\right) \\ DK_0(\theta)^T J\left(\frac{\partial}{\partial\lambda}\nabla H_\lambda^{k_0}(K_0(\theta))\Big|_{\lambda=\lambda_0}\right) \end{pmatrix} = \begin{pmatrix} N_0(\theta)DK_0(\theta)^T J\left(\frac{\partial}{\partial\lambda}\nabla H_\lambda(K_0(\theta))\Big|_{\lambda=\lambda_0}\right) \\ DK_0(\theta)^T J\left(\frac{\partial}{\partial\lambda}\nabla H_\lambda(K_0(\theta))\Big|_{\lambda=\lambda_0}\right) \end{pmatrix} +$$

$$+\begin{pmatrix} N_0(\theta)DK_0(\theta)^T J\left(\frac{\partial}{\partial\lambda}\nabla\left(H_\lambda^{k_0}(K_0(\theta)) - H_\lambda(K_0(\theta)\right)\Big|_{\lambda=\lambda_0}\right) \\ DK_0(\theta)^T J\left(\frac{\partial}{\partial\lambda}\nabla\left(H_\lambda^{k_0}(K_0(\theta)) - H_\lambda(K_0(\theta)\right)\Big|_{\lambda=\lambda_0}\right) \end{pmatrix} = \Lambda_0(\theta) + \Psi_0(\theta).$$

Because $\| \Psi_0 \|_{\rho_0} \le d(\nu+1)|H^{k_0} - H|_{C^3, B_{2r}(K_0)\times A(Q)}$ and $|H^{k_0} - H|_{C^3, B_{2r}(K_0)\times A(Q)} \to 0$ $(k_0 \to \infty)$, for sufficiently large $k_0$,

$$d(\nu+1)|H^{k_0} - H|_{C^3, B_{2r}(K_0)\times A(Q)}\ \tau < \frac{1}{2}$$

holds. We obtain that $|<\Psi_0>| \le d(\nu+1)|H^{k_0} - H|_{C^3, B_{2r}(K_0)\times A(Q)}$. And from $|<\Lambda_0>^{-1}| = \tau_0 < \tau$, $|<\Psi_0(\theta)>| \cdot |<\Lambda_0>^{-1}| < \frac{1}{2}$. Hence $I + <\Lambda_0>^{-1}<\Psi_0>$ is invertible and therefore $<\Lambda_0^{k_0}> = <\Lambda_0>[I + <\Lambda_0>^{-1}<\Psi_0>]$ is invertible. We obtain that

$$<\Lambda_0^{k_0}>^{-1} = (I + <\Lambda_0>^{-1}<\Psi_0>)^{-1} <\Lambda_0>^{-1}.$$

We deform the above equality like that

$$<\Lambda_0^{k_0}>^{-1} = (I + <\Lambda_0>^{-1}<\Psi_0>)^{-1}(I + <\Lambda_0>^{-1}<\Psi_0> - <\Lambda_0>^{-1}<\Psi_0>) <\Lambda_0>^{-1} =$$

$$= <\Lambda_0>^{-1} - (I + <\Lambda_0>^{-1}<\Psi_0>)^{-1} <\Lambda_0>^{-1}<\Psi_0><\Lambda_0>^{-1}.$$

From $|(I + <\Lambda_0>^{-1}<\Psi_0>)^{-1}| \le \sum_{i=0}^{\infty}\left|<\Lambda_0>^{-1}<\Psi_0>\right|^i < 2$, following inequality holds:

$$|<\Lambda_0^{k_0}>^{-1}| \le |<\Lambda_0>^{-1}| + 2d(\nu+1)\tau^2 |H^{k_0} - H|_{C^3, B_{2r}(K_0)\times A(Q)}.$$

Meanwhile from the triangle inequality

$$|H^{k_0}|_{C^3,B_{2r}(K_0)\times A(Q)} \le |H|_{C^3,B_{2r}(K_0)\times A(Q)} + |H^{k_0}-H|_{C^3,B_{2r}(K_0)\times A(Q)}$$

holds.

Because the sum $\sum_{k=1}^{\infty}|H^{k+1}-H^k|_{C^3,B_{2r}(K_0)\times A(Q)}$ converges, for sufficiently large $k_0$, the following inequalities hold:

$$d(\nu+1)|H^k-H|_{C^3,B_{2r}(K_0)\times A(Q)}\,\tau<\frac{1}{4},\ (k\ge k_0), \tag{8}$$

$$|H^k-H|_{C^3,B_{2r}(K_0)\times A(Q)}<1,\ (k\ge k_0), \tag{9}$$

$$2d(\nu+1)\tau^2\left(|H^{k_0}-H|_{C^3,B_{2r}(K_0)\times A(Q)}+\sum_{k=k_0+1}^{\infty}|H^k-H^{k-1}|_{C^3,B_{2r}(K_0)\times A(Q)}\right)<1. \tag{10}$$

Hence

$$|H^{k_0}|_{C^3,B_{2r}(K_0)\times A(Q)}\le|H|_{C^3,B_{2r}(K_0)\times A(Q)}+|H^{k_0}-H|_{C^3,B_{2r}(K_0)\times A(Q)}\le|H|_{C^3,B_{2r}(K_0)\times A(Q)}+1,$$

$$|<\Lambda_0^{k_0}>^{-1}|\le|<\Lambda_0>^{-1}|+2d(\nu+1)\tau^2|H^{k_0}-H|_{C^3,B_{2r}(K_0)\times A(Q)}\le|<\Lambda_0>^{-1}|+1$$

hold and from the definition of $c$

$$c_{k_0}<c. \tag{11}$$

Let's confirm the assumptions (2),(3) in the theorem in section 2 for $e_{k_0}=J\nabla H^{k_0}_{\lambda_0}(K_0(\theta))-\partial_\omega K_0(\theta)$. We obtain that

$$c_{k_0}\gamma^{-4}\delta_0^{-4\sigma}\|e_{k_0}\|_\rho=c_{k_0}\gamma^{-4}\delta_0^{-4\sigma}\|J\nabla H^{k_0}_{\lambda_0}(K_0(\theta))-\partial_\omega K_0(\theta)\|_\rho=$$

$$\le c\gamma^{-4}\delta_0^{-4\sigma}(\|J\nabla H^{k_0}_{\lambda_0}(K_0(\theta))-J\nabla H_{\lambda_0}(K_0(\theta))\|_\rho+\|J\nabla H_{\lambda_0}(K_0(\theta))-\partial_\omega K_0(\theta)\|_\rho)=$$

$$=c\gamma^{-4}\delta_0^{-4\sigma}(\|J\nabla H^{k_0}_{\lambda_0}(K_0(\theta))-J\nabla H_{\lambda_0}(K_0(\theta))\|_\rho+\|e_0\|_\rho)\le$$

$$\le c\gamma^{-4}\delta_0^{-4\sigma}|H^{k_0}-H|_{C^3,B_{2r}(K_0)\times A(Q)}+c\gamma^{-4}\delta_0^{-4\sigma}\|e_0\|_\rho.$$

From the assumption (2) and $|H^{k_0}-H|_{C^3,B_{2r}(K_0)\times A(Q)}\to 0,\ (k_0\to\infty)$, for sufficiently large $k_0$, the following inequality holds:

$$c\gamma^{-4}\delta_0^{-4\sigma}|H^{k_0}-H|_{C^3,B_{2r}(K_0)\times A(Q)}+c\gamma^{-4}\delta_0^{-4\sigma}\|e_0\|_\rho<1.$$

Similarly,

$$c_{k_0}\gamma^{-2}\delta_0^{-2\sigma}\|e_{k_0}\|_\rho\le c\gamma^{-2}\delta_0^{-2\sigma}|H^{k_0}-H|_{C^3,B_{2r}(K_0)\times A(Q)}+c\gamma^{-2}\delta_0^{-2\sigma}|e_0\|_\rho.$$

And from the assumption (3), for sufficiently large $k_0$,

$$c\gamma^{-2}\delta_0^{-2\sigma}|H^{k_0}-H|_{C^3,B_{2r}(K_0)\times A(Q)}+c\gamma^{-2}\delta_0^{-2\sigma}|e_0\|_\rho<r.$$

Thus we obtain that

$$c_{k_0}\gamma^{-4}\delta_0^{-4\sigma}\|e_{k_0}\|_\rho<1, \tag{12}$$

$$c_{k_0}\gamma^{-2}\delta_0^{-2\sigma}\|e_{k_0}\|_\rho<r. \tag{13}$$

Therefore for the pair $(H^{k_0}_{\lambda_0},K_0)$, the assumptions of the theorem in section 2 holds. So there exist $\lambda_{k_0}\in Q$ and a solution $K_{k_0}\in\Pi_{\rho/2}$ of (1) with Hamiltonian $H^{k_0}_{\lambda_0}$ which is non-degenerate. Moreover $\lambda_{k_0}$ and $K_{k_0}$ satisfy

$$\| K_{k_0} - K_0 \|_{\rho/2} \le c\gamma^{-2}\delta_0^{-2\sigma} \| e_0 \|_\rho \le r \,, \tag{14}$$

$$| \lambda_{k_0} - \lambda_0 | \le r \,. \tag{15}$$

Now we take the positive integer number $k_0 \ge 2$ more sufficiently large in order that $k_0$ satisfies

$$A(4^{-(k_0-1)})^{l+2\sigma} \le \| e_0 \|_\rho \,.$$

From (6) for any integer number $k \ge k_0$,

$$| H^k - H^{k-1} |_{C^3, B_{2r}(K_0)\times A(Q)} \le \| e_0 \|_\rho \,(4^{l+2\sigma})^{-k+1} \,. \tag{16}$$

From now on we consider the subsequence $\{H^k\}_{k=k_0}^\infty$ and for simplicity we denote the sequence $\{H^k\}_{k=k_0}^\infty$ as $\{H^k\}_{k=1}^\infty$.

**Second Step.** For the second step, we will prove the following statements:
For any $k \in \mathbf{N}$, there exists $\lambda_k \in Q$ and a map $K_k \in \Pi_{\rho/2^k}$ which is a solution of (1) with Hamiltonian $H_{\lambda_k}^k$. And moreover they satisfy $\| K_k - K_{k-1} \|_{\rho/4^k} \le r(\frac{1}{4^{l+\sigma}})^{k-1}$,

$| \lambda_k - \lambda_{k-1} | \le r(\frac{1}{4^{l+\sigma}})^{k-1}$ $(k \ge 1)$.

We will prove them using induction. We define the following notations:

$$\rho_k = \frac{\rho}{2^{k-1}} \;,\quad \delta_k = \frac{\rho_k}{12} \;,\quad r_k = r(\frac{1}{4^{l+\sigma}})^{k-1},$$

$$B_{r_k}(K_k) = \{ z \in \mathbf{R}^n \,;\, \inf_{\theta \in U_{\rho_k}} | z - K_k(\theta) | < r_k \} \,,$$

$$e_k(\theta) = J\nabla H_{\lambda_k}^k(K_{k-1}(\theta)) - \partial_\omega K_{k-1}(\theta) \,,$$

$$\mu_k = | H^k |_{C^3, B_{r_{k-1}}(K_{k-1})\times Q} \,,\; d_k = \| DK_{k-1} \|_{\rho_{k-1}} \,,\; \nu_k = \| N_{k-1} \|_{\rho_{k-1}} \,,\; \tau_k = | < \Lambda_{k-1}^k >^{-1} | \,.$$

And we define

$$c_k = \lambda(\mu_k,\, d_k,\, \nu_k,\, \tau_k) \,.$$

To prove the statement in second step, we will prove the following five statements with $k \ge 1$, using induction:

A1(k) $| \lambda_k - \lambda_0 | \le r \sum_{i=0}^{k-1} (\frac{1}{4^{l+\sigma}})^i$,

A2(k) $\| K_k - K_0 \|_{\rho_{k+1}} \le r \sum_{i=0}^{k-1} (\frac{1}{4^{l+\sigma}})^i \le r \frac{4^{l+\sigma}}{4^{l+\sigma}-1} \le \frac{4}{3} r$,

A3(k) $c_k \le c$,

A4(k) $c_k \gamma^{-4} \delta_k^{-4\sigma} \| e_k \|_{\rho_k} < 1$,

A5(k) $c_k \gamma^{-2} \delta_k^{-2\sigma} \| e_k \|_{\rho_k} < r_k$.

If A1(k)-A5(k) hold then the assumptions of the theorem 1 in section 2 be satisfied for a pair $(H_{\lambda_k}^k,\; K_{k-1})$. Then we can obtain $\lambda_k$ and $K_k$ from the pair $(H_{\lambda_k}^k,\; K_{k-1})$.

First, consider the case of $k=1$. A1(1) is followed by (15) and A2(1), A3(1), A4(1), A5(1) are respectively followed by (14) , (11), (12), (13) in first step.

Assume that A1(j)-A5(j) hold for $j=1,\cdots,k-1$ and let us prove A1(k)-A5(k). For any $j=2,\cdots,k$, we obtain $K_j$, a solution of (1) with Hamiltonian $H^j_{\lambda_j}$, by applying the theorem 1 in section 2 to $(H^j_{\lambda_j}, K_{j-1})$. Then inequality (4) implies

$$\| K_k - K_0 \|_{\rho_{k+1}} \le \| K_k - K_{k-1} \|_{\rho_{k+1}} + \| K_{k-1} - K_0 \|_{\rho_k} \le$$
$$\le r_k + r\sum_{i=0}^{k-2}\left(\frac{1}{4^{l+\sigma}}\right)^i \le r\sum_{i=0}^{k-1}\left(\frac{1}{4^{l+\sigma}}\right)^i$$

and

$$|\lambda_k - \lambda_0| \le |\lambda_k - \lambda_{k-1}| + |\lambda_{k-1} - \lambda_0| \le$$
$$\le r_k + r\sum_{i=0}^{k-2}\left(\frac{1}{4^{l+\sigma}}\right)^i \le r\sum_{i=0}^{k-1}\left(\frac{1}{4^{l+\sigma}}\right)^i .$$

This proves $A(1k)$ and $A(2k)$.

Let's prove $A(3k)$. From the inequality (9),

$$\mu_k = |H^k|_{C^3,\, B_{r_{k-1}}(K_{k-1})\times Q} \le |H^k|_{C^3, B_{2r}(K_0)\times Q} \le |H|_{C^3, B_{2r}(K_0)\times Q} + |H^k - H|_{C^3, B_{2r}(K_0)\times Q}$$
$$\le |H|_{C^3, B_{2r}(K_0)\times Q} + 1 \le \mu .$$

And from the statement of theorem 1 in section 2, we obtain

$$d_k = \| DK_{k-1} \|_{\rho_{k-1}} \le \| DK_{k-2} \|_{\rho_{k-2}} + \gamma^{-2}\delta_{k-1}^{2\sigma-1}2^{-4\sigma} \le \cdots \le$$
$$\le \| DK_0 \|_{\rho_0} + \gamma^{-2}\delta_{k-1}^{2\sigma-1}2^{-4\sigma} + \cdots + \gamma^{-2}\delta_0^{2\sigma-1}2^{-4\sigma} \le$$
$$\le \| DK_0 \|_{\rho_0} + \sum_{i=0}^{\infty}\gamma^{-2}\delta_i^{2\sigma-1}2^{-4\sigma} \le \| DK_0 \|_{\rho_0} + \gamma^{-2}\delta_0^{2\sigma-1}2^{-4\sigma}\sum_{i=0}^{\infty}(1/2^i)^{2\sigma-1} \le$$
$$\le \| DK_0 \|_{\rho_0} + \gamma^{-2}\delta_0^{2\sigma-1}2^{-4\sigma}\frac{1}{1-2^{-2\sigma+1}} \le \| DK_0 \|_{\rho_0} + \gamma^{-2}\delta_0^{2\sigma-1}\frac{1}{2^{4\sigma}-2^{2\sigma+1}} \le$$
$$\le d_0 + \beta = d .$$

Similarly, we get

$$\nu_k = \| N_{k-1} \|_{\rho_{k-1}} \le \nu_0 + \beta = \nu .$$

Now let us prove the following inequalities by using induction:

$$\tau_k = |<\Lambda_{k-1}^{H_k}>^{-1}| \le \tau_0 + \beta + 1 = \tau ,$$

$$|<\Lambda_{k-1}^{H_{k-1}}>^{-1}| \le \tau_0 + \beta + 1 = \tau .$$

In the case of $k=1$ is trivial. Assume that they hold for $i=0,\cdots,k-1$.

Performing similar computation in step 1, we obtain

$$\Lambda_{k-1}^{k}(\theta)=\begin{pmatrix} N_{k-1}(\theta)DK_{k-1}(\theta)^T J\left(\frac{\partial}{\partial\lambda}\nabla H_{\lambda}^{k-1}(K_{k-1}(\theta))\Big|_{\lambda=\lambda_{k-1}}\right) \\ DK_{k-1}(\theta)^T J\left(\frac{\partial}{\partial\lambda}\nabla H_{\lambda}^{k-1}(K_{k-1}(\theta))\Big|_{\lambda=\lambda_{k-1}}\right)\end{pmatrix}+$$

$$+\begin{pmatrix} N_{k-1}(\theta)DK_{k-1}(\theta)^T J\left(\frac{\partial}{\partial\lambda}\nabla\left(H_{\lambda}^{k}(K_{k-1}(\theta))-H_{\lambda}^{k-1}(K_{k-1}(\theta)\right)\Big|_{\lambda=\lambda_{k-1}}\right) \\ DK_{k-1}(\theta)^T J\left(\frac{\partial}{\partial\lambda}\nabla\left(H_{\lambda}^{k}(K_{k-1}(\theta))-H_{\lambda}^{k-1}(K_{k-1}(\theta)\right)\Big|_{\lambda=\lambda_{k-1}}\right)\end{pmatrix}=$$

$$=\Lambda_{k-1}^{k-1}(\theta)+\Psi_{k-1}(\theta).$$

In this place we get

$$\|\Psi_{k-1}(\theta)\|_{\rho_{k-1}}\le d_k(\nu_k+1)\,|H^k-H^{k-1}|_{C^3,\,B_{\eta_{k-1}}(K_{k-1})\times Q}\le d(\nu+1)\,|H^k-H^{k-1}|_{C^3,\,B_{2r}(K_0)\times Q}\,.$$

From (8), we obtain

$$d(\nu+1)\,|H^k-H^{k-1}|_{C^3,B_{2r}(K_0)\times Q}\cdot|<\Lambda_{k-1}^{k-1}>^{-1}|<d(\nu+1)\,|H^k-H^{k-1}|_{C^3,B_{2r}(K_0)\times Q}\cdot\tau\le\frac{1}{2}\,.$$

Thus

$$|<\Psi_{k-1}>|\cdot|<\Lambda_{k-1}^{k-1}>^{-1}|\le\frac{1}{2}\,.$$

Hence $I+<\Lambda_{k-1}^{k-1}>^{-1}<\Psi_{k-1}>$ is invertible and therefore

$$<\Lambda_{k-1}^{k}>=<\Lambda_{k-1}^{k-1}>[I+<\Lambda_{k-1}^{k-1}>^{-1}<\Psi_{k-1}>]$$

is invertible, i.e.

$$<\Lambda_{k-1}^{k}>^{-1}=[I+<\Lambda_{k-1}^{k-1}>^{-1}<\Psi_{k-1}>]^{-1}<\Lambda_{k-1}^{k-1}>^{-1}.$$

We deform of expression of $<\Lambda_{k-1}^{k}>^{-1}$ like that

$$<\Lambda_{k-1}^{k}>^{-1}=(I+<\Lambda_{k-1}^{k-1}>^{-1}<\Psi_{k-1}>)^{-1}[I+<\Lambda_{k-1}^{k-1}>^{-1}<\Psi_{k-1}>-<\Lambda_{k-1}^{k-1}>^{-1}<\Psi_{k-1}>]<\Lambda_{k-1}^{k-1}>^{-1}=$$
$$=<\Lambda_{k-1}^{k-1}>^{-1}-[I+<\Lambda_{k-1}^{k-1}>^{-1}<\Psi_{k-1}>]^{-1}<\Lambda_{k-1}^{k-1}>^{-1}<\Psi_{k-1}><\Lambda_{k-1}^{k-1}>^{-1}\,.$$

Since $|(I+<\Lambda_{k-1}^{k-1}>^{-1}<\Psi_{k-1}>)^{-1}|\le\sum_{i=0}^{\infty}|<\Lambda_{k-1}^{k-1}>^{-1}<\Psi_{k-1}>|^{i}<2$, we get

$$|<\Lambda_{k-1}^{k}>^{-1}|\le|<\Lambda_{k-1}^{k-1}>^{-1}|+2d(\nu+1)\tau^2\,|H^k-H^{k-1}|_{C^3,B_{2r}(K_0)\times Q} \qquad (17).$$

And from the theorem 1 in section 2, we get

$$|<\Lambda_{k-1}^{k-1}>^{-1}|<|<\Lambda_{k-2}^{k-1}>^{-1}|+\gamma^{-2}\delta_{k-1}{}^{2\sigma-1}2^{-4\sigma} \qquad (18).$$

Appling (17),(18) repeatedly, we obtain

$$|<\Lambda_{k-1}^{k}>^{-1}|\le|<\Lambda_{k-2}^{k-1}>^{-1}|+2d(\nu+1)\tau^2\,|H^k-H^{k-1}|_{C^3,B_{2r}(K_0)\times Q}+\gamma^{-2}\delta_{k-1}{}^{2\sigma-1}2^{-4\sigma}\le\cdots$$

$$\cdots\le|<\Lambda_0>^{-1}|+\sum_{j=1}^{k}2d(\nu+1)\tau^2\,|H^j-H^{j-1}|_{C^3,B_{2r}(K_0)\times Q}+\sum_{j=1}^{k}\gamma^{-2}\delta_{j-1}{}^{2\sigma-1}2^{-4\sigma}\le$$

$$\le\ |<\Lambda_0>^{-1}|+\sum_{j=1}^{\infty}2d(\nu+1)\tau^2\,|H^j-H^{j-1}|_{C^3,B_{2r}(K_0)\times Q}+\sum_{j=1}^{\infty}\gamma^{-2}\delta_{j-1}{}^{2\sigma-1}2^{-4\sigma}\le$$

$$\le|<\Lambda_0>^{-1}|+1+\beta=\tau\,.$$

(here, we used (10)). Therefore we obtain

$$|<\Lambda_{k-1}^{k-1}>^{-1}|\le|<\Lambda_0>^{-1}|+1+\beta=\tau\,.$$

Hence for any $k\in\mathbf{N}$, $\tau_k=|<\Lambda_{k-1}^{k}>^{-1}|\le\tau$ holds. From definition of $c$, we get $c_k\le c$. This implies A3(k).

Now let's prove $A(4k)$. We get

$$|H^k-H^{k-1}|_{C^3,B_{2r}(K_0)\times Q}\le\|e_0\|_{\rho_0}\,(4^{l+2\sigma})^{-k+1}$$

from (16). Performing some computation, we obtain that

$$c_k\gamma^{-4}\delta_k{}^{-4\sigma}\,\|e_k\|_{\rho_k}=c\gamma^{-4}(\delta_0/2^{k-1})^{-4\sigma}\,\|J\nabla H_{\lambda_{k-1}}^{k}(K_{k-1}(\theta))-\partial_\omega K_{k-1}(\theta)\|_{\rho_k}=$$
$$\le c\gamma^{-4}(\delta_0/2^{k-1})^{-4\sigma}(\|J\nabla H_{\lambda_{k-1}}^{k}(K_{k-1}(\theta))-J\nabla H_{\lambda_{k-1}}^{k-1}(K_{k-1}(\theta))\|_{\rho_k}+$$
$$+\|J\nabla H_{\lambda_{k-1}}^{k-1}(K_{k-1}(\theta))-\partial_\omega K_{k-1}(\theta)\|_{\rho_k})=$$
$$=c\gamma^{-4}\delta_0{}^{-4\sigma}(2^{k-1})^{4\sigma}\,\|J\nabla H_{\lambda_{k-1}}^{k}(K_{k-1}(\theta))-J\nabla H_{\lambda_{k-1}}^{k-1}(K_{k-1}(\theta))\|_{\rho_k}\le$$
$$\le c\gamma^{-4}\delta_0{}^{-4\sigma}\,|H^k-H^{k-1}|_{C^3,B_{2r}(K_0)\times Q}\,(2^{k-1})^{4\sigma}\le$$
$$\le c\gamma^{-4}\delta_0{}^{-4\sigma}\,\|e_0\|_{\rho_0}\,(4^{l+2\sigma})^{-k+1}(2^{k-1})^{4\sigma}\le$$
$$\le(4^{l+2\sigma})^{-k+1}(4^{k-1})^{2\sigma}\le(\frac{1}{4^l})^{k-1}\le 1.$$

Therefore A4(k) holds. Similarly,

$$c_k\gamma^{-2}\delta_k{}^{-2\sigma}\,\|e_k\|_{\rho_k}=c\gamma^{-2}(\delta_0/2^{k-1})^{-2\sigma}\,\|J\nabla H_{\lambda_{k-1}}^{k}(K_{k-1}(\theta))-\partial_\omega K_{k-1}(\theta)\|_{\rho_k}=$$
$$\le c\gamma^{-2}(\delta_0/2^{k-1})^{-2\sigma}(\|J\nabla H_{\lambda_{k-1}}^{k}(K_{k-1}(\theta))-J\nabla H_{\lambda_{k-1}}^{k-1}(K_{k-1}(\theta))\|_{\rho_k}+$$
$$+\|J\nabla H_{\lambda_{k-1}}^{k-1}(K_{k-1}(\theta))-\partial_\omega K_{k-1}(\theta)\|_{\rho_k})=$$
$$=c\gamma^{-2}\delta_0{}^{-2\sigma}(2^{k-1})^{2\sigma}\,\|J\nabla H_{\lambda_{k-1}}^{k}(K_{k-1}(\theta))-J\nabla H_{\lambda_{k-1}}^{k-1}(K_{k-1}(\theta))\|_{\rho_k}\le$$
$$\le c\gamma^{-2}\delta_0{}^{-2\sigma}\,|H^k-H^{k-1}|_{C^3,B_{2r}(K_0)\times Q}\,(2^{k-1})^{2\sigma}\le$$
$$\le c\gamma^{-2}\delta_0{}^{-2\sigma}\,\|e_0\|_{\rho_0}\,(4^{l+2\sigma})^{-k+1}(2^{k-1})^{2\sigma}\le$$
$$\le r(4^{l+2\sigma})^{-k+1}(4^{k-1})^{\sigma}\le r(\frac{1}{4^{l+\sigma}})^{k-1}=r_k$$

and this implies A5(k). Hence A1(k)-A5(k) hold for all $k\in\mathbf{N}$.

Let's $\lambda_k$ and $K_k$ are the parameter and the map obtained by applying the theorem in section 2 $k$-times. Then $\lambda_k$ and $K_k$ satisfies the following inequality from theorem 1 in section 2:

$$\| K_k - K_{k-1} \|_{\rho/4^k} \le \| K_k - K_{k-1} \|_{\rho_k} \le r_k = r\left(\frac{1}{4^{l+\sigma}}\right)^{k-1},$$

$$| \lambda_k - \lambda_{k-1} | < r_k = r(4^{-l+\sigma})^{k-1}.$$

Therefore $\{\lambda_k\}$ converges to a certain parameter $\lambda_\infty \in Q$. And from lemma 4, the sequence of real analytic maps $\{K_k\}$ converges to certain map $K_\infty \in C^1(\mathbf{T}^n, \mathbf{U})$. Because for any $k \in \mathbf{N}$, $K_k$ satisfy (1) with Hamiltonian $H^k_{\lambda_k}$ and $H_k$ $C^3$-converges to $H$ in $\mathbf{T}^n$, therefore $K_\infty$ is a solution of (1) with $H_{\lambda_\infty}$. □

## References

[Arnold 1963] Arnold, V.I.: *Proof of a Theorem of A.N.Kolmogorov on the Invariance of Quasi-Periodic Motions under Small Perturbations of the Hamiltonian*, Russian Mathematical Surveys 18(9), (1963) 9-36. in VLADIMIR I. ARNOLD Collected Works1, Springer, 2009, 267-294.

[Broer et al. 1990] Broer, H.W., Huitema G.B., Takens F., Braaksma B.L.J.: *Unfoldings and bifurcations of quasi-periodic tori*, Mem. Amer. Math. Soc. **83** (421) (1990), i–vii and 1–175.

[Broer- Sevryuk 2008] Broer H. W., Sevryuk M. B.; *KAM Theory: quasi-periodicity in dynamical systems*, Handbook of Dynamical Systems vol 3, ed Broer H. W. et al (Amsterdam: North-Holland) at press, 2008, 1-136.

[Broer et al. 2007] Broer, H.W., Hoo J., Naudot V.: *Normal linear stability of quasi-periodic tori*, J. Differential Equations **232** (2007), 355–418.

[Casartelli 1983] Casartelli, M.: *Relaxation Times and Randomness for a Nonlinear Classical System,* Lecture Notes in Physics Vol.179, Springer, (1983), 252-253.

[Celletti-Chierchia 2006] Celletti, A., Chierchia L.: *KAM Stability for a three-body problem of the Solar system*, Z. angew. Math. Phys. 57 (2006) 33–41.

[Ciocci et al. 2005] Ciocci, M.-C., Litvak-Hinenzon A., Broer H.W.: *Survey on dissipative* KAM *theory including quasi-periodic bifurcation theory, based on lectures by Henk Broer*, GeometricMechanics and Symmetry. The Peyresq Lectures (Peyresq, 2000–01), J. Montaldi and T.S. Ratiu, eds, London Math. Soc. Lecture Note Series, Vol. 306, Cambridge Univ. Press, Cambridge (2005), 303–355.

[Fèjoz 2004] Fèjoz, J.: *Démonstration du « théor ème d'Arnold » sur la stabilité du système planétaire (d'après M. Herman)*, Ergod. Th. & Dynam. Sys. 24 (2004), 1-62.

[Gidea, 2009] Gidea, M., Meiss J. D., Ugarcovicic I., Weiss H.: *Applications of KAM Theory to Population Dynamics*, Journal of Biological Dynamics, (2009), 1–23.

[Haro-Llave 2004a] Haro, A., de la Llave R.; *A parameterization method for the computation of whiskers in quasi periodic maps: rigorous results*, Preprint mp_arc 04-348, 2004.

[Haro-Llave 2004b] Haro, A., de la Llave R.; *A parameterization method for the computation of whiskers in quasi periodic maps: numerical implementation*, Preprint mp_arc 04-350, 2004.

[Helleman- Kheifets 1985] Helleman, R., Kheifets S. A.: *Nonlinear Dynamics Aspects of Modern Storage Rings*, Lecture Notes in Physics 247, Springer, (1985), 64-76.

[Herman 1986] Herman, M.R.: *Sur les courbes invariantes par les difféomorphismes de l'anneau*, Vol. 1 and 2, Astérisque 103–104 (1983), i and 1–221; 144 (1986), 1–248.

[Hoo 2005] Hoo, J.; *Quasi-periodic bifurcations in a strong resonance: Combination tones in gyroscopic stabilisation*, Ph.D. Thesis, Univ. Groningen (2005).

[Huitema 1988] Huitema, G.B.: *Unfoldings of quasi-periodic tori*, Ph.D. Thesis, Univ. Groningen (1988).

[Jong-Paek 2012] Jong , Wu-Hwan ; Paek , Jin-Chol ; *Existence of Invariant Tori for Differentiable Hamiltonian Vector Field without Action-Angle Variables*, arXiv.org/math-ph/1208.2083, pp.1-12, 2012.

[Kolmogorov 1954] Kolmogorov, A.N.: *Preservation of Conditionally Periodic Movements with Small Change in the Hamilton Function* (Russian), Akad. Nauk. S.S.S.R., Doklady 98 (1954) 527-535. English: Los Alames Scientific Laboratory translation LA-TR-71-67 by Helen Dahl. Reprinted in: G. Casati and J. Ford eds: Stochastic Behavior in Classical and Quantum Hamiltonian Systems, Lecture Notes in Physics Vol. 93, Springer, 1979, 51-56.

[Li-Yi 2002] Li, Yong; Yi, Yingfei: *Persistence of invariant tori in generalized Hamiltonian systems*, Ergod. Th. & Dynam. Sys. 22 (2002), 1233–1261.

[Llave et al. 2005] Llave, R. de la, González, A., Jorba, À., Villanueva J.: *KAM theory without action-angle variables*, Nonlinearity 18 (2005), 855–895.

[Locatelli-Giorgilli 2007] Locatelli, U., Giorgilli, A.: *Invariant tori in the Sun-Jupiter-Saturn System*, AIM science.org Discrete and Continuous Dynamical Systems-Sires B, Vol. 7, No 2, March (2007), 377-398.

[Lorentz 1986] Lorentz, G.G.: *Bernstein Polynomials 2ed*, Chelsea Publishing Company, 1986.

[Moser 1962] Moser, J.K.: *On invariant curves of area-preserving mappings of an annulus*, Nach. Akad. Wiss. Göttingen, Math. Phys. Kl. II 1 (1962), 1-20.

[Moser 1966] Moser, J.K.: *A rapidly convergent iteration method and non-linear differential equations*, I, Annali della Scuola Norm. Super, de Pisa ser. III, 20, (1966), 265-315; II, (1966), 499-535.

[Rüssmann 1970] Rüssmann, H.: Kleine Nenner I: *Über invariante Kurven differenzierbarer Abbildungen eines Kreisringes*, Nachr. Akad.Wiss. Göttingen,Math.-Phys. Kl. II 5 (1970), 67–105.

[Takens 1971] Takens, F.: *A* $C^1$ *counterexample to Moser's twist theorem*, Indag. Math. 33 (1971), 378–386.